# Dendritic flux jumps in an organic superconducting crystal

M. M. Mola[*] and R. Haddad
*Department of Physics & Physical Science, Humboldt State University, Arcata, CA 95501*

S. Hill
*Department of Physics, University of Florida, Gainesville, Florida 32611-8440*

**Abstract**

Angle dependant torque magnetization measurements have been carried out on the organic superconductor, $\kappa$-$(ET)_2Cu(NCS)_2$ at extremely low temperatures (25 - 300 mK). Magneto-thermal instabilities are observed in the form of abrupt magnetization (flux) jumps for magnetic field sweeps of 0 – 20 Tesla. A fractal analysis of the flux jumps indicate that the instabilities do show a self similar structure with a fractal dimension of varying between 1.15 – 1.6. The fractal structure of the flux jumps in our sample shows a striking similarity to that of $MgB_2$ thin film samples, in which magneto-optical experiments have recently shown that the small flux jumps are due to the formation of dendritic flux structures. These smaller instabilities act to suppress the critical current density of the thin films. The similarity of the flux jump structure of our samples suggests that we are also observing the dendritic instability, but in a bulk sample rather than a thin film. This is the first observation of the dendritic instability in a bulk superconducting sample, and is likely due to the layered nature of $\kappa$-$(ET)_2Cu(NCS)_2$, which results in a quasi-two dimensional flux structure over the majority it's mixed state phase diagram.

## 1. Introduction

Low dimensionality plays a prominent roll in contemporary condensed matter physics, with its importance highly apparent when investigating vortex interactions within a type-II superconductor due to an applied magnetic field. Thin film samples are often used to investigate two-dimensional (2D) vortices, and have resulted in the observation of dendritic instabilities as well as conventional flux jumps in $MgB_2$,[1-2] $Nb$[3] and $YBa_2Cu_3O_7$[4] thin films. However, the dendritic instability has not yet been observed in a bulk crystalline superconductor, as it is generally believed that large vortex-vortex repulsions must dominate over the rigidity of the three dimensional (3D) flux lattice.[5] Likewise the sample must be thermally isolated such that the temperature distribution within the sample remains highly non-uniform during dendritic growth.[6] One material which provides both of these characteristics at extremely low temperatures is the layered organic superconductor $\kappa$-$(ET)_2Cu(NCS)_2$, where ET stands for bis-ethlenedithio-tetrathiafulvalene.

Previous work on $\kappa$-$(ET)_2Cu(NCS)_2$, has shown that the majority of the mixed state phase diagram is dominated by the quasi-two-dimensional (Q2D) vortex solid and the pancake vortex liquid.[7-8] In fact, the decoupling transition from the 3D flux lattice to the Q2D vortex solid, which consists of nearly uncorrelated layers of pancake vortex lattices, occurs at an essentially temperature independent applied magnetic field strength of ~ 7 mT.[9] Thus, above this field, each layer of the superconductor behaves almost completely independently of its neighboring layers, essentially giving a stack of autonomous thin films, each one unit cell thick. Moreover, at extremely low temperatures (~ 25 – 300 mK), the melting of the Q2D vortex lattice has been attributed to quantum fluctuations.[7] However, for applied magnetic fields below that required to melt the flux lattice, magneto-thermal instabilities, in the form of flux jumps have been observed in our previous torque magnetization measurements.[7, 10-11] In most bulk crystalline samples, the observed flux jumps are large, indicating that a sizable portion of the sample participates in the magneto-thermal instability. In contrast, vortex avalanches in $\kappa$-(ET)$_2$Cu(NCS)$_2$ are observed for both the increasing and decreasing field sweeps (see figure 1), resulting in a decrease in the critical current density of the sample, and the size of the flux jumps cover the entire spectrum of sizes, from $\Delta M \sim 10^{-3}M$, up to macroscopic instabilities involving a large portion of the sample. Figure 1 indicates that the presence of small flux jumps (SFJ) is a temperature dependent phenomenon. At the lowest temperatures (25 mK, see figure 2), the majority of the flux jumps are small and irregular. However, as the temperature is increased, the size of the flux jumps increases until they eventually evolve into the much more regular conventional flux jumps (CFJ) one would expect for a bulk sample (110 mK). As the temperature increases further still, the CFJ cease, leaving the expected smooth magnetization curve for a type-II superconductor in the mixed state.

It has previously been suggested that the magneto-thermal instabilities in this material result from the thermal isolation of the crystal due to Kapitza resistance, essentially a phonon mismatch between the superconducting crystal and the surrounding cryogenic fluids.[12] Thus, no instabilities are observed above a temperature of approximately 200 mK, above which the crystal can maintain thermal contact with the surrounding cryogens. In figure 1, the observed flux jumps show both large CFJ as well as smaller, irregular flux jumps, which have been associated with dendritic instabilities in thin film samples.[13] Moreover, by examining the flux jumps at a continuously finer scale, we notice that the jumps seem to have a "statistically self-similar" structure. In other words, the structure of the flux jumps appears the same regardless of the scale of magnification (up to a point). As self similarity defines fractal behavior, we shall extend the analysis of our previous work to show that the flux jumps observed in $\kappa$-(ET)$_2$Cu(NCS)$_2$ exhibit a rich and complex structure that has not before been observed in a bulk crystalline material.[11] We will therefore apply a fractal analysis to the magnetization as a function of applied magnetic field at various temperatures and angles and show that these plots have curves with fractal dimensions.

## 2. Experimental

High quality single crystals were grown using standard techniques.[14] Samples had approximate dimensions of 1.0 × 1.0 × 0.3 mm$^3$ and were mounted on a capacitive cantilever beam torque magnetometer, which was attached to a single axis rotator. This allowed for a variation of the angle of $\theta$ between the **a**-axis of the crystal and the applied magnetic field from 0° to 90°. The entire assembly was then loaded directly into the mixing chamber of a top loading $^3$He/$^4$He dilution refrigerator situated in the bore of a 20 tesla superconducting magnet at the Nation High Magnetic Field Laboratory (NHMFL). The applied field was swept at a constant rate of 0.5 T/min. for all measurements. Angle dependent measurements were preformed at 25 mK, at approximately 5° intervals between 16° and 88° between the **a**-axis of the crystal and the applied field. Temperature dependent measurements were preformed at temperatures of 25, 60, 87, 110, 130, 150, 200 and 300 mK, at an angle of θ = 47°.

To determine the fractal dimension of the observed flux jumps, we employed a "box counting" algorithm as described is reference 14. This algorithm consists of laying a grid across the magnetization versus field plots and then counting the number of boxes required to completely enclose the M vs. B plot. Once the number of boxes is recorded, the scale of the applied grid is decreased and the number of grid boxes is again counted, which continues to ever decreasing box size. When one plots log(N) verse log(1/A), where N is the number of boxes at each iteration and A is the area of each box, the slope of this plot gives the fractal "box counting" dimension.[15] The box counting dimension gives a pragmatic approximation to the "Housdorf" dimension. To ensure that our algorithm worked, we first applied it to several well known mathematical fractals (i.e. the Cantor set and the Koch curve[15]). Our analysis of the Cantor set gave a box counting dimension of $D_B = 0.635 \pm 0.001$, where the similarity dimension is $D_S = \frac{\log(2)}{\log(3)} = 0.631$, giving a 0.6% error. Likewise the box counting dimension of the Koch curve was determined to be $D_B = 1.284 \pm 0.002$, where the similarity dimension is $D_S = \frac{\log(4)}{\log(3)} = 1.262$, giving an error of less than 2%. The resulting error in both of these examples is likely due to the finite nature of the data sets used to create the Cantor set and the Koch curve. In building the prefractals, each iteration resulted in an exponential increase in the number of data points compared to the previous iteration. Thus, after only 10 iterations the data sets making up the prefractals were incredibly large and difficult to manipulate. However, even with only 10 iterations, the error was reduced to less than 2%. Thus, we felt confident that our box counting program could accurately determine the fractal dimension of the magnetization versus magnetic field data obtained for this organic superconductor.

### 3. Results

In figure 1, we plot the magnetization versus applied magnetic field at several different temperatures where the angle between the applied field and the **a**-axis of the crystal is 47°. Clearly the structure of the flux jumps is different at high temperatures than at low temperatures, where SFJ are observed at temperatures of 25 and 60 mK. At higher temperatures SFJ no longer observed; instead, semi-regular CFJ which involve a large portion of the sample in the instability (i.e., ΔM ~ 1/3M) are observed. In fact, at the highest measured temperatures (~300 mK), flux jumps are no longer observed as the crystal is now able to transfer thermal energy to its surroundings at a rate such that the instabilities no longer occur.

In figure 2, the upsweep and down-sweep (indicated by arrows) of the magnetization versus applied magnetic field are plotted for a sample at 25 mK and an angle of 47°. Both the upsweep and down sweep show similar structure with conventional flux jumps (CFJ) mixed with a large number of smaller irregular flux jumps. At this temperature, all angles except 88° show similar behavior.[16] In the insets to figure 2, magnifications of several regions of the upsweep are plotted. At low applied field strength (both on the upsweep and the down-sweep), the flux jumps are regular and of a large size, consistent with conventional flux jumps, as seen in many other bulk materials, including this one.[1-4, 12] Following a CFJ, the system recovers and the magnetization tends toward the smooth envelope which defines the critical current density at a given applied field and temperature. However, upon closer inspection of the magnetization at larger applied magnetic fields, the flux jumps no longer show a regular pattern, but are irregular in size and frequency, and the magnetization no longer attempts to recover to the main magnetization curve due to the high frequency of the instabilities. This suppression of the critical current density by the SFJ has been associated with the rapid growth of dendritic avalanches in thin film samples of many different superconductors.[1-4, 13]

We believe that the similar structure of magnetization versus applied magnetic field in $\kappa$-$(ET)_2Cu(NCS)_2$, as compared to thin film samples of various other superconductors, is the direct result of dendritic avalanches of magnetic flux into the interior of our crystal. Both the temperature dependence of the flux jumps, and the low threshold field at which SFJs are observed, corroborate this assertion.[1 & 13] However, whereas the instabilities within the thin film samples are directly observed,[1] we must infer the dendritic structures from the similarities between the magnetization of the bulk sample and that of the thin films.[13] Our results now beg the question, why would a behavior associated with 2D thin films be observed in a bulk sample. We believe the similarity between the two stems from the Q2D nature of $\kappa$-$(ET)_2Cu(NCS)_2$ within the mixed state. As stated earlier, we know that the majority of the mixed state phase diagram is dominated by the Q2D flux solid and pancake liquid phases. Thus, the high anisotropy of this material results in the decoupling of adjacent layers, leading the system to resemble a stack of independent 2D sheets. The resulting avalanches of magnetic flux during a field swept experiment consists of a large number of pancake vortices within a given layer jumping into the crystal interior along dendritic channels, which are weakly correlated to the avalanches in adjacent layers. These Q2D instabilities suppress the critical current density within the superconducting state due to the local heating effects of migratory flux. The small size of the flux jumps is a result of the dendrites in each layer. Perhaps it is not surprising that a phenomenon seen previously only in thin films is now observed within a highly anisotropic organic superconductor ($\gamma \sim 200$ [ref.14]) which exhibits a dimensional crossover at low applied fields.

To determine the box counting (fractal) dimension of the magnetization versus applied magnetic field for the down-sweep data shown in figure 2 we have plotted log(N) versus log(1/A) in figure 3. There are three distinct regions in the plot. The first is at the very bottom of the plot where the area of each box is large and therefore the number of boxes is few. Thus, when plotted, the data appears quantized in steps. In the second region we actually obtain the fractal dimension from the slope of the plot. For this particular angle and temperature ($47^o$ & 25 mK) the fractal dimension is $1.509 \pm 0.006$, clearly a non-integer dimension. The third region of the plot also displays a linear slope, however, in this region the box size is so small that the one data point per box limit is approached and the slope is close to zero.[15]

When similar analyses are carried out on data at other angles at this same temperature, the fractal dimensions range in value from 1.15 to 1.6. There seems to be little correlation between the fractal dimension and the angle $\theta$. However, the dimension of the down-sweep is typically larger than that of the upsweep. Analysis of the temperature dependence of the fractal dimension shows that, as the temperature increases, the dimension moves closer to unity as one would expect for a smooth line. Again, the fractal dimension of the magnetization is another indicator that the flux structure is due to an underlying complex mechanism, such as dendritic avalanches of magnetic flux.

## 4. Conclusion

Conventional as well as small flux jumps have been observed in the organic superconductor $\kappa$-(ET)$_2$Cu(NCS)$_2$. Using a simple box counting algorithm we have determined the fractal dimension of the magnetization versus applied magnetic field for the coldest temperatures (25 mK), and at various angles between the applied field and the **a**-axis of the crystal. Moreover, we have observed a magnetization signature which has been shown to be due to the dendritic migration of magnetic flux in to the interior of thin film samples of MgB$_2$ and various other superconducting films. Thus, we believe that we are also observing a dendritic instability in a bulk crystal which is likely due to the quasi-two-dimensional behavior of this highly anisotropic material. It is likely that this instability would only be seen in highly anisotropic materials ($\gamma \sim 200$) which are also relatively clean, as is the case for the material in this study (i.e. there are very few defects in the organic superconductors compared to the high temperature superconductors; see ref. 14).

## 5. Acknowledgments

Work carried out at the NHMFL was supported by a cooperative agreement between the State of Florida and NSF under DMR-9527035.

**Figure Captions**

Fig. 1. Magnetization versus applied magnetic field at an angle $\theta = 47^{o}$ between the **a**-axis and the B-field (the plots are offset for clarity). At the lowest temperature, both CFJ and SFJ are observed, where the SFJ act to significantly decrease the critical current density. At higher temperatures the flux jumps become more regular (CFJ) and at the highest temperatures they cease altogether.

Fig. 2. Magnetization versus applied magnetic field at an angle $\theta = 47^{o}$ and a temperature of 25 mK. Upon closer inspection of the magnetization curve, one finds increasing levels of detail which look statistically self similar. Self similarity is the signature of fractal structure.

Fig. 3. To determine the fractal dimension of the magnetization versus applied field plot from figure 2, we must take the slope of the above figure. It results in a box counting dimension of $1.509 \pm 0.006$, a non-integer value.

Figure 1

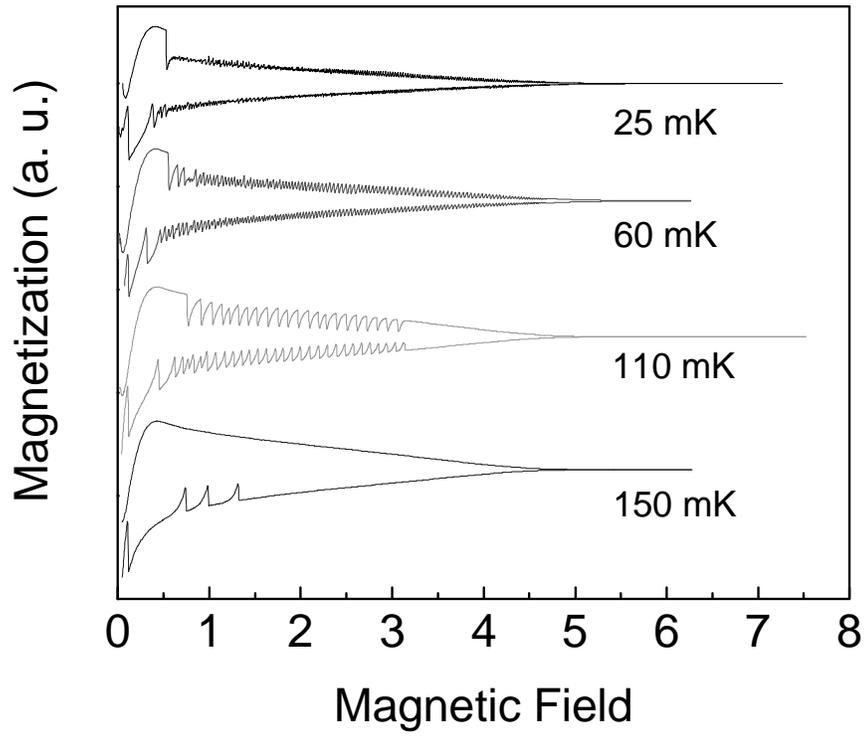

Figure 2

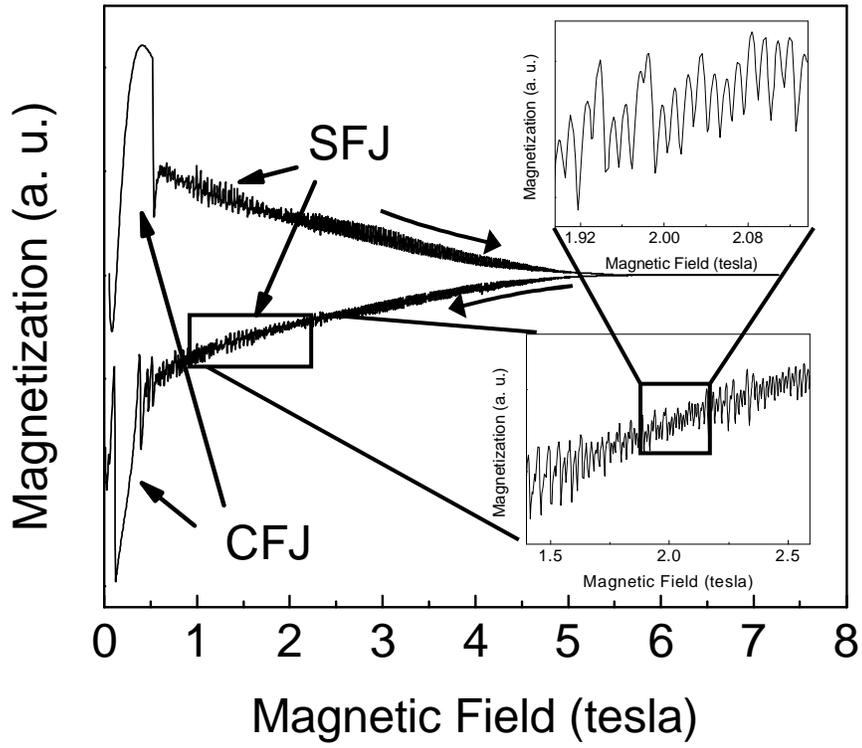

Figure 3

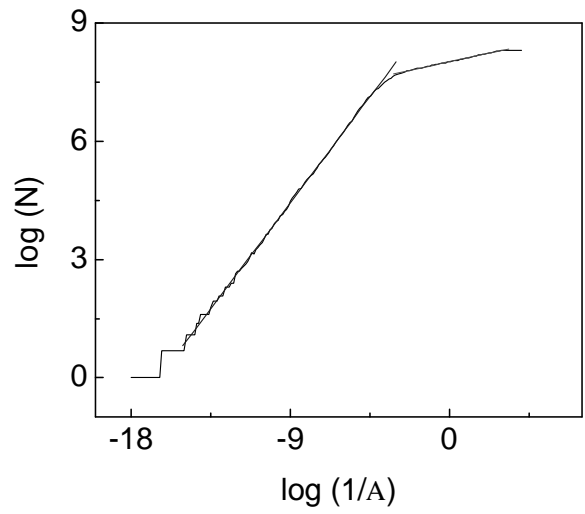

# 6. References

\*   Email address: mmm67@humboldt.edu

1.  T. H. Johansen, M. Baziljevich, D. V. Shantsev, P. E. Goa, Y. M. Galperin, W. N. Kang, H. J. Kim, E. M. Choi, M. –S. Kim, S. I. Lee, Europhysics Letters, **59**, 599 (2002).

2.  Z. W. Zhao, S. L. Li, Y. M. Ni, H. P. Yang, Z. Y. Liu, H. H. Wen, W. N. Kang, H. J. Kim, E. M. Choi, S. I. Lee, Physical Review B, **65**, 064512 (2002).

3.  C. A. Duran, P. L. Gammel, R. E. Miller, D. J. Bishop, Physical Review B, **52**, 75 (1995).

4.  P. Leiderer, J. Boneberg, P. Brull, V. Bujok, S. Herminghaus, Physical Review Letters, **71**, 2646 (1993).

5.  G. Blatter, M. V. Feigel'man, V. B. Geshkenbein, A. I. Larkin, V. M. Vinokur, Reviews of Modern Physics, **66**, 1125 (1994).

6.  I. Aranson, A. Gurevich, V. Vinokur, Physical Review Letters, **87**, 067003 (2001).

7.  M. M. Mola, S. Hill, J. S. Brooks, J. S. Qualls, Physical Review Letters, **86**, 2130 (2001).

8.  M. M. Mola, J. T. King, C. P. McRaven, S. Hill, J. S. Qualls, J. S. Brooks, Physical Review B, **62**, 5965 (2000).

9.  S. L. Lee, F. L. Pratt, S. J. Blundell, C. M. Aegerter, P. A. Pattenden, K. H. Chow, E. M. Forgan, T. Sasaki, W. Hayes, H. Keller, Physical Review Letters, **79**, 1563 (1997).

10. S. Hill, M. M. Mola, J. S. Qualls, J. S. Brooks, Synthetic Metals, **133-134**, 221 (2003).

11. M. M. Mola, S. Hill, J. S. Qualls, J. S. Brooks, International Journal of Modern Physics B, **15**, 3353 (2001).

12. A. G. Swanson, J. S. Brooks, H. Anzai, N. Konoshita, M. Tokumoto, K. Murata, Solid State Communications, **73**, 353-356 (1990).

13. F. L. Barkov, D. V. Shantsev, T. H. Johansen, P.E. Goa, W. N. Kang, H. J. Kim, E. M. Choi, S. I. Lee, Physical Review B, **67**, 064513 (2003).

14. T. Ishiguro, K. Yamaji, and G. Saito, *Organic Superconductors*, Springer-Verlag, Berlin, 1998.

15. P. S. Addison, *Fractals and Chaos: An Illustrated Course*, IOP Publishing, Bristol and Philadelphia, 1997.

16. $88^{\circ}$ between the **a**-axis and the magnetic field is within the angle for a lock-in transition in which the magnetic flux gets "locked-in" between the layers of the superconducting material, forming Josephson vortices.